\documentclass[twocolumn,aps,prl,superscriptaddress]{revtex4}
\usepackage{color}
\usepackage{amsmath}
\usepackage{amssymb}
\usepackage{graphicx}

\makeatletter

\usepackage{epsfig}
\usepackage{color}
\usepackage{ulem}

\makeatother

\begin{document}
\title{Auxiliary-Bath Numerical Renormalization Group Method and Successive Collective Screening in Multi-Impurity Kondo Systems}
\author{Danqing Hu}
\affiliation{Beijing National Laboratory for Condensed Matter Physics and Institute
of Physics, Chinese Academy of Sciences, Beijing 100190, China}
\affiliation{Department of Physics and Chongqing Key Laboratory for Strongly Coupled Physics, Chongqing University, Chongqing 401331, China}
\author{Jiangfan Wang}
\affiliation{School of Physics, Hangzhou Normal University,  Hangzhou, Zhejiang 311121, China}
\author{Yi-feng Yang}
\email[]{yifeng@iphy.ac.cn}
\affiliation{Beijing National Laboratory for Condensed Matter Physics and Institute
of Physics, Chinese Academy of Sciences, Beijing 100190, China}
\affiliation{School of Physical Sciences, University of Chinese Academy of Sciences, Beijing 100049, China}
\affiliation{Songshan Lake Materials Laboratory, Dongguan, Guangdong 523808, China}

\date{\today}

\begin{abstract}
We propose an auxiliary-bath algorithm for the numerical renormalization group (NRG) method to solve multi-impurity models with shared electron baths. The method allows us to disentangle the electron baths into independent Wilson chains to perform standard NRG procedures beyond the widely adopted independent bath approximation. Its application to the 2-impurity model immediately reproduces the well-known even- and odd-parity channels. For 3-impurity Kondo models, we find collective screening of cluster degrees of freedom depending on impurity configurations and clarify the false prediction of a non-Fermi liquid ground state for the $C_3$ symmetric case in previous literature due to improper treatment of disentanglement. Our work highlights the importance of nonlocal spatial correlations due to shared baths and reveals a generic picture of successive collective screening for entropy depletion that is crucial in real correlated systems. Our method greatly expands the applicability of the NRG and opens an avenue for its further development.
\end{abstract}

\maketitle

Quantum impurity models are of fundamental importance for correlated electron physics and have practical applications in mesoscopic systems \cite{Sasaki2000Nature,Jamneala2001,Park2002Nature,Pustilnik2004JPCM, Bork2011NatPhys, Pruser2014NatComm, Keller2015Nature, Iftikhar2018Science, Knorr2022PRL}. As minimal interacting models, they exhibit many exotic quantum many-body phenomena such as the local Fermi liquid  \cite{Kondo1964,Anderson1961,Nozieres1974,Hewson1993}, non-Fermi liquid (NFL) with anyon excitations \cite{Tsvelick1985, Pang1991,Zarand2000,Yotsuhashi2002,Lopes2020,Lotem2022}, and the competition between Kondo screening and the induced Rudermann-Kittel-Kasuya-Yosida (RKKY) interaction that underlies the basic physics of heavy fermion systems \cite{Ruderman1954,Kasuya1956,Yosida1957,Doniach1977,Colemanbook}. Quantum impurity models also have been used to understand lattice problems and material properties within the framework of the dynamical mean-field theory \cite{Rozenberg1994,Georges1996, Maier2005}.

Among all methods developed for solving the impurity models \cite{Wilson1975, Bulla2008,Caffarel1994, Kajueterthesis1996,Hirsch1986,Gull2011}, Wilson's numerical renormalization group (NRG) \cite{Wilson1975,Bulla2008} has been successfully applied to the single-impurity Kondo problem and can in principle yield highly accurate results over wide energy and temperature ranges for very general impurity interactions. However, its application to multi-impurity models has been severely limited by its own methodology, which requires to first map the electron baths to conduction (Wilson) chains coupled to the impurities \cite{Ferrero2007, Zitko2007, Zitko2008, Mitchell2011, Sela2011, Bork2011, Mitchell2012,Mitchell2013, Eickhoff2020,Wojcik2020PRB}. This prerequisite is generally not possible except for very rare cases \cite{Jones1987, Jones1988, Zhu2011, Mitchell2015,Paul1996, Paulthesis2000,Ingersent2005}, because the impurities often couple to the shared baths in different momentum-dependent form that cannot be easily disentangled. Therefore, many works assume independent baths with artificial exchange interactions between impurities, based on which various phase diagrams have been predicted \cite{Ferrero2007,Zitko2007, Zitko2008, Mitchell2011, Sela2011,Bork2011,Mitchell2012, Mitchell2013,Eickhoff2020,Wojcik2020PRB,Konig2021}, but real systems often have shared baths that not only cause the Kondo screening but also mediate the RKKY interaction. This hinders broader applications of the NRG.

In this work, we propose a general strategy named the auxiliary-bath NRG to disentangle the shared baths into a number of independent auxiliary baths that can be mapped to the Wilson chains following the standard NRG procedure. Our method naturally reproduces the effective odd- and even-parity channels for 2-impurity model \cite{Zhu2011}, and reveals collective screening of cluster degrees of freedom for 3-impurity models depending on configuration details \cite{Wang2024}, where the RKKY interactions are produced automatically without artificial manipulation. We find no sign of NFL ground state predicted in previous literatures for the $C_3$ symmetric model \cite{Paul1996,Paulthesis2000,Ingersent2005}. Our work highlights the importance of shared-bath models  in real systems, and opens an avenue for further development of the NRG method.

We consider as an example the $N$-impurity model consisting of two parts, $H=H_{\text{I}}+H_{\text{B}}$, where $H_{\text{I}}$ represents a general model of interacting impurities ($f_{\mu\sigma}$) and $H_{\text{B}}$ describes their coupling to a free electron bath ($c_{\boldsymbol{k}\sigma}$):
\begin{equation}
H_{\text{B}}=\sum_{\boldsymbol{k}\sigma}\epsilon_{\boldsymbol{k}}c_{\boldsymbol{k}\sigma}^{\dagger}c_{\boldsymbol{k}\sigma}+\sum_{\boldsymbol{k}\mu\sigma}\left(V_{\boldsymbol{k}\mu}f_{\mu\sigma}^{\dagger}c_{\boldsymbol{k}\sigma}+\text{H.c.}\right) 
\end{equation}
where $V_{\boldsymbol{k}\mu}=V_{\mu}e^{i\boldsymbol{k\cdot r_{\mu}}}$ with $V_\mu$ being the local hybridization strength of the $\mu$-th ($\mu =1,\cdots, N$) impurity with the bath at the lattice site $\boldsymbol{r}_{\mu}$. The shared-bath model is generally unsolvable for traditional NRG algorithm because it is impossible to disentangle the bath and generate the Wilson chain for each impurity independently. This is easy to see because independent chains cannot induce the inter-impurity RKKY interaction. 

To overcome this difficulty, we notice that as long as only the impurity degrees of freedom are considered, we may integrate out the bath electrons and derive the effective action: 
\begin{equation}
S^{\text{eff}}_{\text{I}} =S_{{\rm I}}-\sum_{\mu\nu\sigma n}f_{\mu\sigma,n}^{\dagger}\mathcal{B}_{\mu\nu}(i\omega_{n})f_{\nu\sigma,n},
\end{equation}
where $S_{{\rm I}}$ is the action of the impurities alone given by $H_{\text{I}}$, $\mathcal{B}_{\mu\nu}(i\omega_{n})=\sum_{\boldsymbol{k}}\frac{V_{\mu\boldsymbol{k}}V_{\nu\boldsymbol{k}}^{*}}{i\omega_{n}-\epsilon_{\boldsymbol{k}}}$ is induced by their coupling to the shared bath, and the subscript $n$ denotes the fermionic Matsubara frequency $\omega_{n}$. Hence, the solution only requires correct $\mathcal{B}_{\mu\nu}(i\omega_{n})$ regardless of its origin. This motivates us to introduce an effective model with auxiliary baths coupled $\boldsymbol{k}$-independently to all impurities: 
\begin{equation}
H_{\text{AU}}=\sum_{\boldsymbol{k}p\sigma}E_{\boldsymbol{k}}^p\tilde{c}_{\boldsymbol{k}p\sigma}^{\dagger}\tilde{c}_{\boldsymbol{k}p\sigma}+\sum_{\boldsymbol{k}\mu p\sigma}(W_{\mu p}f_{\mu\sigma}^{\dagger}\tilde{c}_{\boldsymbol{k}p\sigma}+\text{H.c.}), 
\end{equation}
where $\tilde{c}_{\boldsymbol{k}p\sigma}$ ($p=0,1,\cdots,N_{\text{A}}-1$) denote the $N_A$ auxiliary baths satisfying
\begin{equation}
\sum_{p\boldsymbol{k}}\frac{W_{\mu p}W_{\nu p}^{*}}{i\omega_{n}-E_{\boldsymbol{k}}^{p}}=\sum_{\boldsymbol{k}}\frac{V_{\mu\boldsymbol{k}}V_{\nu\boldsymbol{k}}^{*}}{i\omega_{n}-\epsilon_{\boldsymbol{k}}}=\mathcal{B}_{\mu\nu}(i\omega_{n}).
\label{eq:5}
\end{equation}
Replacing the sum over momentum by integration over energy, we derive the equations:
\begin{equation}
\sum_{p}w_{\mu p}w_{\nu p}^{*}\tilde{\rho}_{p}(\omega)=\rho_{\mu\nu}(\omega)\equiv \sum_{\boldsymbol{k}}e^{i\boldsymbol{k\cdot}(\boldsymbol{r}_{\mu}-\boldsymbol{r}_{\nu})}\delta(\omega-\epsilon_{\boldsymbol{k}}),
\label{eq:6}
\end{equation}
where we have defined the dimensionless coupling parameters $w_{\mu p}=W_{\mu p}/V_\mu$, and $\tilde{\rho}_{p}(\omega)$ are the densities of states of the auxiliary baths to be solved for NRG. We have the diagonal term $\rho_{\mu\mu}(\omega)=\rho_0(\omega)\equiv\sum_{\boldsymbol{k}}\delta(\omega-\epsilon_{\boldsymbol{k}})$, which is the density of states (DOS) of the original bath. Integrating over the energy leads to $\sum_{p}w_{\mu p}w^*_{\nu p}=\delta_{\mu\nu}$, which implies that $\{w_{\mu p}\}$ forms an $N\times N_A$ complex matrix whose row vectors $\boldsymbol{w}_\mu\equiv\{w_{\mu p}\}$ are orthonormalized. Physically, this follows from the mapping $c_{\mu\sigma} \rightarrow \sum_{\boldsymbol{k}p}w_{\mu p}\tilde{c}_{\boldsymbol{k}p\sigma}=\sum_{p}w_{\mu p}\tilde{c}_{p\sigma}$ and the anticommutation relation. The fact that each auxiliary bath couples to all impurities $f_{\mu\sigma}$ in a momentum-independent manner in $H_{\text{AU}}$ allows the standard NRG procedure to transform it into an independent Wilson chain.

\begin{figure}[t]
\begin{centering}
\includegraphics[width=0.46\textwidth]{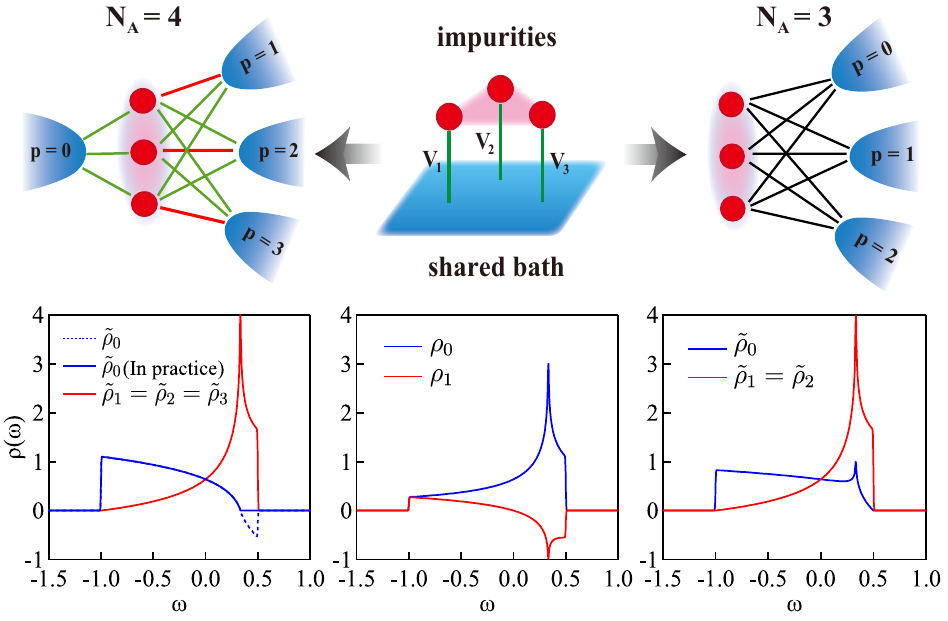}
\par\end{centering}
\caption{Illustration of different mapping strategies that disentangle the original shared bath to $N_A=3$ and 4 auxiliary baths for the $C_3$ symmetric 3-impurity model on a triangular lattice. The green (red) lines for $N_A=4$ indicate a positive (negative) sign of the dimensionless coupling parameter $w_{\mu p}$. Also compared are the densities of states (DOS) of the original bath and the auxiliary baths. The narrow negative $\tilde{\rho}_0(\omega)$ region (dashed line) of the auxiliary baths for $N_A=4$ is replaced by zero in practice \cite{supp}. The bandwidth of the original bath is set to $D=1.5$ as the basic energy scale.}
\label{fig1}
\end{figure}

In general, $\rho_{\mu\nu}(\omega)$ for $\mu\neq\nu$ are complex so that $w_{\mu p}$ should also be complex. However, if $\epsilon_{\boldsymbol{k}}=\epsilon_{-\boldsymbol{k}}$,  e.g., under spatial inversion or time reversal symmetry, we have $\text{Im}\rho_{\mu\nu}(\omega)=\sum_{\boldsymbol{k}}\sin[{\boldsymbol{k}\boldsymbol{\cdot}(\boldsymbol{r}_\mu-\boldsymbol{r}_\nu)}]\delta(\omega-\epsilon_{\boldsymbol{k}})=0$, and $w_{\mu p}$ may take real values. Since $\rho_{\mu\nu}=\rho_{\nu\mu}^*$, the minimal number of independent equations and auxiliary baths to reproduce all $\rho_{\mu\nu}(\omega)$ for general configurations should be $N_A=\frac{N(N-1)}{2}+1$, and all spatial information of impurities is absorbed in the DOS $\tilde{\rho}_{p}(\omega)$ of these auxiliary baths. For instance, the 2-impurity model may have a solution with $N_{\text{A}}=2$ and solving Eq. (\ref{eq:6}) gives $\tilde{\rho}_{p}(\omega)=\rho_0(\omega)\pm \rho_{12}(\omega)$ and $\boldsymbol{w}_{\mu}=\{1/\sqrt{2},\pm1/\sqrt{2}\}$, which are exactly the effective baths coupled to the odd- and even-parity combinations of local orbitals in previous work \cite{Zhu2011}. {For 3-impurity models, we prove that the minimal number of auxiliary baths is $N_A=4$ for general configurations and $3$ for the $C_3$ symmetric case \cite{supp}. For general $N$-impurity models, a systematic strategy to construct auxiliary baths is given in Ref. \cite{Wang2024}.

Obviously, the above procedure is independent of the impurity Hamiltonian $H_\text{I}$. To show the advantage of our approach, we apply it to the 3-impurity model of different configurations on a triangular lattice, which is a minimal model for exploring the interplay of Kondo physics and magnetic frustrations but has not been thoroughly understood \cite{Paul1996, Paulthesis2000,Ingersent2005,Kudasov2002, Savkin2005, Lazarovits2005, Wang3IK2022}. To compare with the literature \cite{Paul1996}, we take $H_{\text{I}}=U\sum_\mu (n_{\mu\uparrow}^f-\frac{1}{2})(n_{\mu\downarrow}^f-\frac{1}{2})$ and perform calculations on the effective spin model after projecting out the impurity charge degrees of freedom via the Schrieffer-Wolff (SW) transformation \cite{supp}:
\begin{equation}
H=\sum_{\boldsymbol{k}p\sigma}E_{\boldsymbol{k}}^p\tilde{c}_{\boldsymbol{k}p\sigma}^{\dagger}\tilde{c}_{\boldsymbol{k}p\sigma}+\sum_{pp^{\prime}\mu}J_{\mu}^{pp^{\prime}}\boldsymbol{S}_{\mu}\cdot\boldsymbol{s}_{pp'},
\label{eq:Hs}
\end{equation}
in which $J_{\mu}^{pp^{\prime}}=\frac{8V^2}{U}w_{\mu p}^*w_{\mu p^{\prime}}$, $\boldsymbol{S}_\mu=\frac{1}{2}\sum_{\alpha\beta}f_{\mu\alpha}^{\dagger}\boldsymbol{\sigma}_{\alpha\beta} f_{\mu\beta}$, and $\boldsymbol{s}_{pp'}=\frac{1}{2}\sum_{\alpha\beta}\tilde{c}_{p\alpha}^{\dagger}\boldsymbol{\sigma}_{\alpha\beta}\tilde{c}_{p'\beta}$. For simplicity, we  set $V_\mu=V$ for all three impurities. Since the mapping operation commutes with the SW transformation, the above model can also be obtained by applying the mapping directly to the Kondo Hamiltonian using $\boldsymbol{s}_{\mu} \rightarrow \sum_{pp'}w^*_{\mu p}w_{\mu p'}\boldsymbol{s}_{pp'}$, which generates Kondo scattering between auxiliary baths that is missing under independent bath assumption. One can easily prove that the RKKY interactions derived from Eq. (\ref{eq:Hs}) is identical to those from the original model \cite{supp}. 

\textit{$C_3$ symmetric model.---} {For clarity, we set $|\boldsymbol{r}_\mu-\boldsymbol{r}_\nu|=a$ ($a$ is the lattice parameter) and  $\rho_{\mu\nu}(\omega)=\rho_1(\omega)$ for $\mu\neq \nu$. We first apply the $N_A=3$ mapping with $w_{\mu p}=\frac{1}{\sqrt{3}}e^{-i2\pi(\mu-1)p/3}$ ($p=0,1,2$), $\tilde{\rho}_0(\omega)=\rho_0(\omega)+2\rho_1(\omega)$, and $\tilde{\rho}_1(\omega)=\tilde{\rho}_2(\omega)=\rho_0(\omega )-\rho_1(\omega)$ as derived in the supplemental material \cite{supp} and plotted in Fig. \ref{fig1}. The effective model is then solved for different hybridization strength $\alpha=\pi (V/D)^2$ using standard NRG, with the discretization parameter $\Lambda=10$ \cite{units}, $N_s=3000$ kept states, and $z$-averaging over $z=0$, 1/4, 1/2, 3/4 \cite{supp}.
 
\begin{figure}[t]
\begin{centering}
\includegraphics[width=0.49\textwidth]{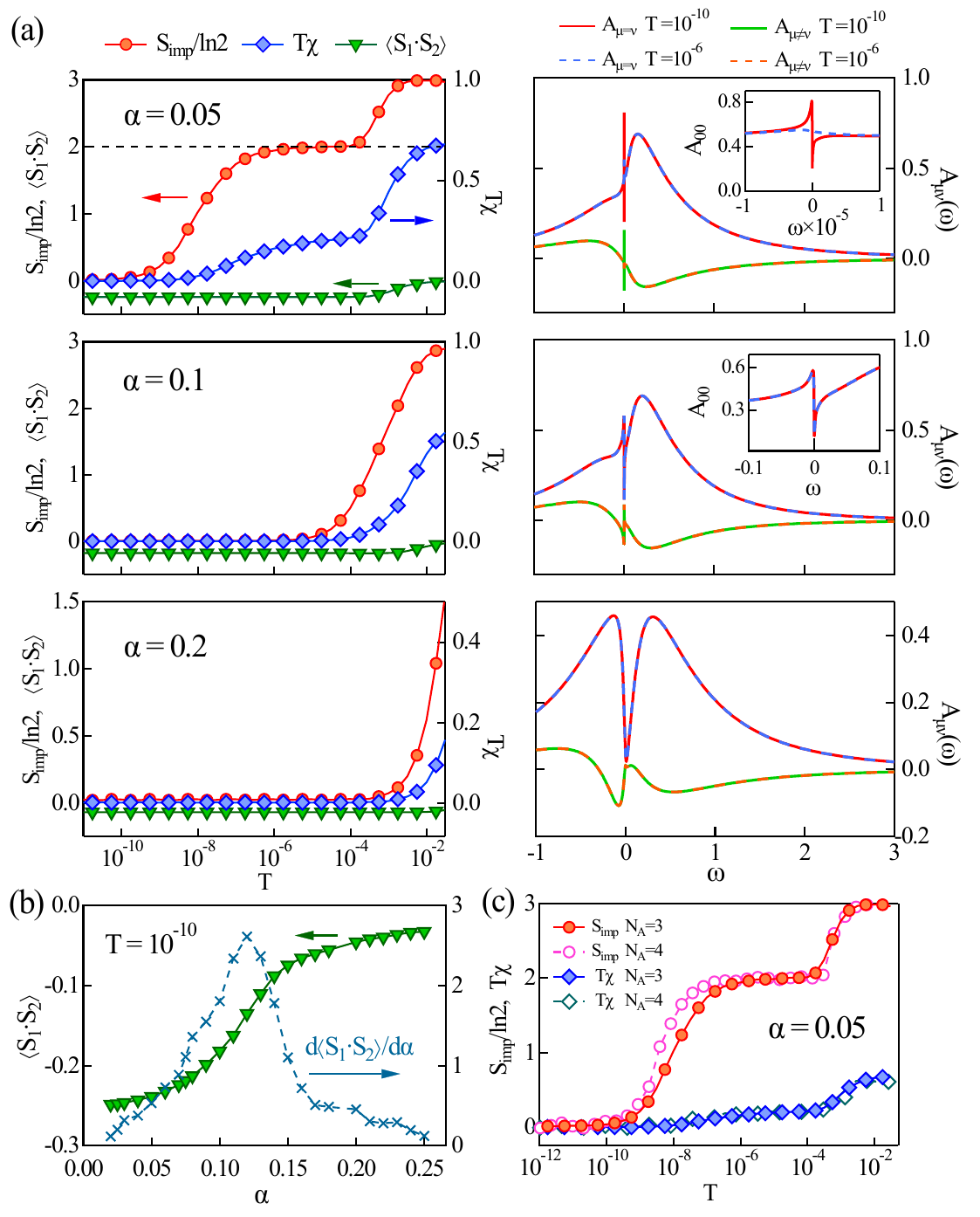}
\par\end{centering}
\caption{Auxiliary-bath NRG results for the 3-impurity model with $C_3$ symmetry. (a) Comparison of the impurity entropy $S_{{\rm imp}}$, the spin susceptibility $\mu_{\rm eff}\equiv T\chi$, and the spin correlation function $\langle\boldsymbol{S}_{1}\boldsymbol{\cdot S}_{2}\rangle$ as functions of the temperature $T$ for $\alpha=0.05$, 0.1, 0.2. Also presented are the local ($\mu=\nu$) and nonlocal ($\mu\neq\nu$) spectra of the original bath at $T=10^{-10}$ and $10^{-6}$. The insets show the features around zero energy in an enlarged plot. (b) Evolution of $\langle\boldsymbol{S}_{1}\boldsymbol{\cdot S}_{2}\rangle$ and its derivative as a function of $\alpha$ at $T=10^{-10}$. (c) Comparison of the entropy and spin susceptibility for $\alpha=0.05$ using two different mapping strategies with $N_A=3$ and 4.}
\label{fig2} 
\end{figure}

Figure \ref{fig2}(a) plots the results of the impurity entropy $S_{{\rm imp}}=S-S_0$, the spin correlation function $\langle\boldsymbol{S}_{1}\boldsymbol{\cdot S}_{2}\rangle$, the spin susceptibility, $\mu_{\rm eff}=T\chi=\langle S_z^2\rangle - \langle S_z^2 \rangle_0$, where $S$ is the entropy and $S_z$ the $z$-component spin of the whole system. The subscript $0$ refers to noninteracting systems without impurities \cite{Zitko2006}. Also calculated are the spectra, $A_{\mu\nu}(\omega)=-\frac{1}{\pi} \text{Im}G_{\mu\nu}(\omega)$, where $G_{\mu\nu}(\omega)$ is the retarded Green's function of the original bath \cite{supp}. At high temperatures, the entropy always approaches $3\ln2$ and $\mu_{\rm eff}=T\chi=3/4$, indicating free local moments without inter-impurity correlations. With lowering temperature, $\langle\boldsymbol{S}_{1}\boldsymbol{\cdot S}_{2}\rangle$ turns negative, marking the onset of antiferromagnetic (AFM) RKKY correlations between impurities. Accordingly, the entropy decreases and, for $\alpha=0.05$, first drops to $2\ln2$ as $\langle\boldsymbol{S}_{1}\boldsymbol{\cdot S}_{2}\rangle$ approaches $-1/4$ and $\mu_{\rm eff}$ reduces to $1/4$. This arises from the four-fold degenerate cluster impurity states characterized by the total spin $S^{\rm tot}=1/2$ and helicity $h=\pm1$ \cite{supp}. The $S^{\rm tot}=3/2$ high spin quartet with $h=0$ have a higher energy and are quenched in this temperature region. At even lower temperatures, both $S_{\rm imp}$ and $\mu_\text{eff}$ drop to zero, indicating collective Kondo screening of the cluster spin degrees of freedom. For larger $\alpha=0.1$ and 0.2, the entropy drops directly from $3\ln 2$ to 0, below which $\mu_{\rm eff}$ remains zero, implying that the spins are quickly screened under strong Kondo coupling. However, as shown in Fig. \ref{fig2}(b), the spin correlation function increases gradually from -1/4 to a small negative value as $\alpha$ increases, indicating the destruction of cluster spin degrees of freedom and the crossover from collective screening of cluster spin states to individual screening of each single spin state.Correspondingly, $A_{\mu\nu}(\omega)$ at $\alpha=0.05$ develop a sharp dip-peak structure around zero energy in the screened state ($T=10^{-10}$) compared to the unscreened state ($T=10^{-6}$), which gradually turns into a pseudogap in the local ($\mu=\nu$) spectra and an asymmetric lineshape in the nonlocal ($\mu\neq\nu$) spectra at larger $\alpha$.

The $N_A=3$ results provide a benchmark for the mapping strategy with $N_A=4$, where the coupling parameters satisfy $|w_{\mu p}|=1/2$ with signs illustrated by different colors  in Fig. \ref{fig1} \cite{supp}. The effective model contains one auxiliary bath $\tilde{\rho}_0(\omega)=\rho_0(\omega)+3\rho_1(\omega)$ coupled equally to all impurities and three others ($p=1,2,3$) of the same $\tilde{\rho}_p(\omega)=\rho_0(\omega)-\rho_1(\omega)$ coupled differently. One may notice that $\tilde{\rho}_0(\omega)$ is negative in a narrow high energy window (dashed line), which has no influence on low energy physics and can be safely set to zero for approximation \cite{supp}. We solve the model by using the interleaving approach \cite{Mitchell2014,Stadler2016}, keeping 3000 states after adding the $p=0$ bath and 20000-50000 states after adding other three together \cite{supp}. Figure \ref{fig2}(c) compares the calculated $S_{{\rm imp}}$ and $\mu_{\rm eff}$ at $\alpha=0.05$ for $N_A=3$ and $4$, confirming the good consistency between two mapping strategies.

Previous works have predicted a NFL state with irrational degeneracy for the $C_3$ symmetric 3-impurity model \cite{Paul1996,Paulthesis2000,Ingersent2005,Lazarovits2005}. Here we find no sign of such exotic state. To clarify the origin of this discrepancy, we rewrite our effective Hamiltonian in their proposed form \cite{supp}. The corresponding parameters are constrained in a region of independent Kondo screening in the phase diagram proposed in Ref. \cite{Paul1996}, suggesting that the predicted frustrated NFL fixed point \cite{Paul1996,Paulthesis2000,Ingersent2005} is an artefact of their independent parameter assumption, which ignores falsely the parameter relations in the effective model and pushes the phase space into unphysical regions. The NFL was also obtained by perturbative renormalization group calculations on a similar effective Hamiltonian in Ref. \cite{Lazarovits2005}, possibly due to incorrect treatment of disentanglement. It will be interesting to see if such exotic fixed point might be realized in other circumstances.

\begin{figure}[t]
\includegraphics[width=0.48\textwidth]{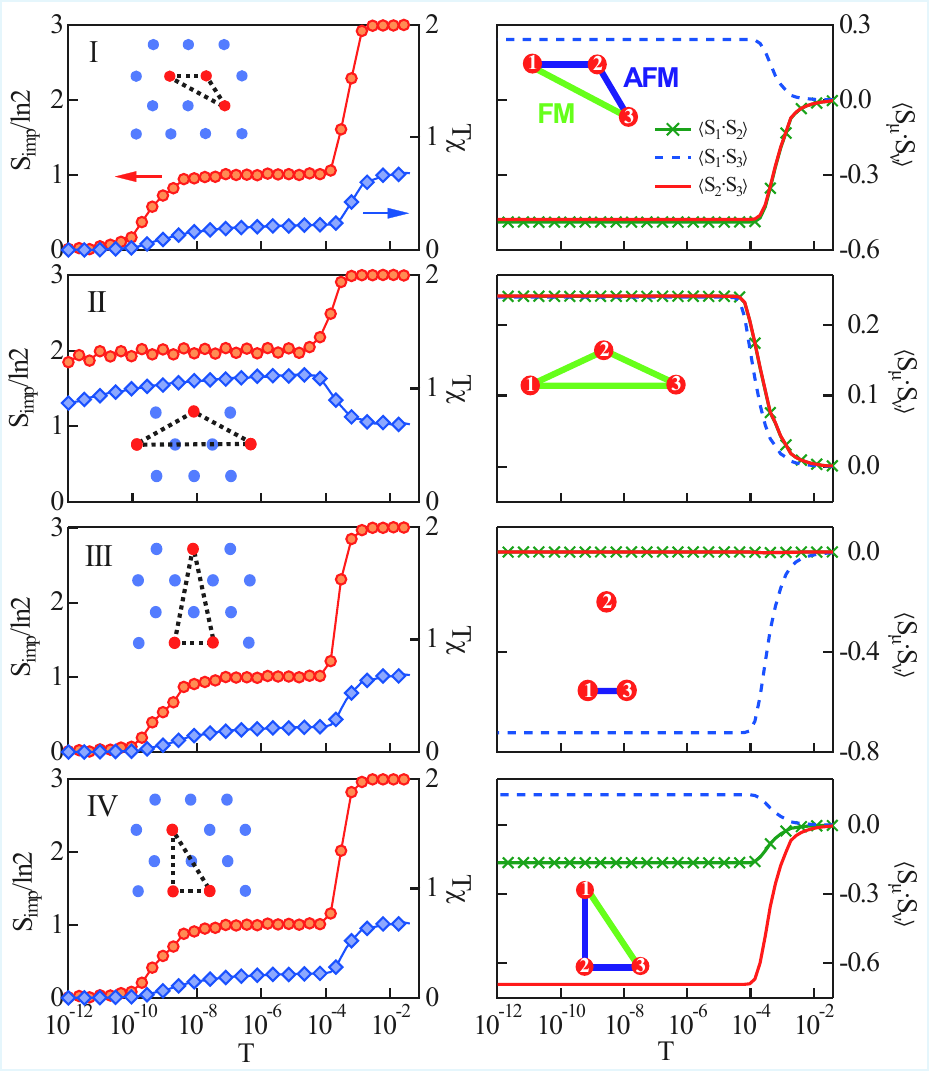}
\caption{}{Auxiliary-bath NRG results for 3-impurity models of four typical configurations without $C_3$ symmetry, showing the impurity entropy $S_{\rm imp}$, the spin susceptibility $\mu_\text{eff}$, and the spin correlation functions $\langle\boldsymbol{S}_\mu\cdot \boldsymbol{S}_\nu\rangle$ as functions of the temperature at a fixed $\alpha=0.05$. The insets show the impurity configurations and the induced ferromagnetic (FM) or antiferromagnetic (AFM) correlations between impurities.}
\label{fig3}
\end{figure}

\textit{General 3-impurity models.---}We now apply the $N_A=4$ mapping to configurations without $C_3$ symmetry. Figure \ref{fig3} shows the results for $\alpha=0.05$ using the same NRG parameters with $z$-averaging over $z=0$ and 1/2.

Configuration I takes an obtuse isosceles triangle form. The emergent RKKY interactions cause ferromagnetic (FM) correlation between $\boldsymbol{S}_1$ and $\boldsymbol{S}_3$, and AFM correlation between $\boldsymbol{S}_2$ and both others. At intermediate temperatures, the entropy $S_{\rm imp}$ reduces from $3\ln2$ to $\ln2$ and $\mu_\text{eff}$ drops to 1/4 (one third of its free value), suggesting an emergent spin-1/2 doublet with dominant AFM correlations between impurities of shortest distance. With further lowering temperature, both the entropy and spin susceptibility approach zero while the spin correlation functions remain unchanged, indicating collective screening of the cluster degree of freedom. 

Configuration II has a similar triangle form but of larger size. The correlations between impurities become all FM, marking the formation of a spin-3/2 quartet demonstrated by the entropy plateau at $2\ln 2$ and the enhanced $\mu_{\rm eff}$ at intermediate temperatures. Both quantities start to decrease at the lowest temperature,  marking the onset of collective screening of the quartet.

Configuration III forms an acute isosceles triangle. We find AFM correlations between impurities 1 and 3 on neighboring sites, which form a spin-singlet bond decoupled almost completely from impurity 2 with vanishing $\langle\boldsymbol{S}_{1}\boldsymbol{\cdot S}_{2}\rangle$ and $\langle\boldsymbol{S}_{2}\boldsymbol{\cdot S}_{3}\rangle$. Correspondingly, $S_{{\rm imp}}$ decreases from $3\ln2$ to $\ln2$ and $\mu_{\rm eff}$ reduces to 1/4, as expected for the decoupled spin-1/2 impurity, which is eventually screened as both quantities approach zero at lower temperatures. The singlet bond, however, exhibits constant $\langle\boldsymbol{S}_{1}\boldsymbol{\cdot S}_{3}\rangle$ and persists down to the lowest temperature due to the small $\alpha$.

Configuration IV is a right-angled triangle that exhibits similar inter-impurity correlations as configuration I. The impurities also form an effective spin-1/2 cluster, giving rise to the $\ln2$ entropy plateau and reduced $\mu_{\rm eff}=1/4$. Further lowering the temperature causes collective screening of the cluster degree of freedom and depletes all the impurity entropy.

Overall, our method captures automatically the different cluster spin states with emergent RKKY interactions for general impurity configurations, and confirms the presence of collective screening of these cluster degrees of freedom over wide parameter regions.

\textit{Discussions and conclusion.---}
We have proposed an auxiliary-bath strategy to extend NRG to multi-impurity models with shared electron baths. Our method provides an unambiguous way to disentangle the baths with two important consequences: (1) It allows for inter-bath scattering that may help destabilize a non-Fermi liquid ground state; (2) It correctly produces the RKKY interaction and avoids artificial treatment of inter-impurity correlations. The mapping also gives the auxiliary parameters without uncontrolled approximations. 

Its application to 3-impurity models reveals important observations summarized schematically in Fig. \ref{fig4}. As the hybridization increases or the temperature decreases, the impurities evolve first from free local moments to some emergent cluster spin states depending on the configurations. As the Kondo effect sets in, the cluster degrees of freedom may first be screened such that, over a wide  parameter region, magnetic correlations between impurities remain strong even though the magnetic entropy is fully quenched. Such collective screening of cluster degrees of freedom reflects the most fundamental physics in multi-impurity Kondo systems, while the usual local Kondo screening of each impurity individually occurs only when the hybridization strength is sufficiently strong.

\begin{figure}[t]
	\begin{centering}
		\includegraphics[width=0.48\textwidth]{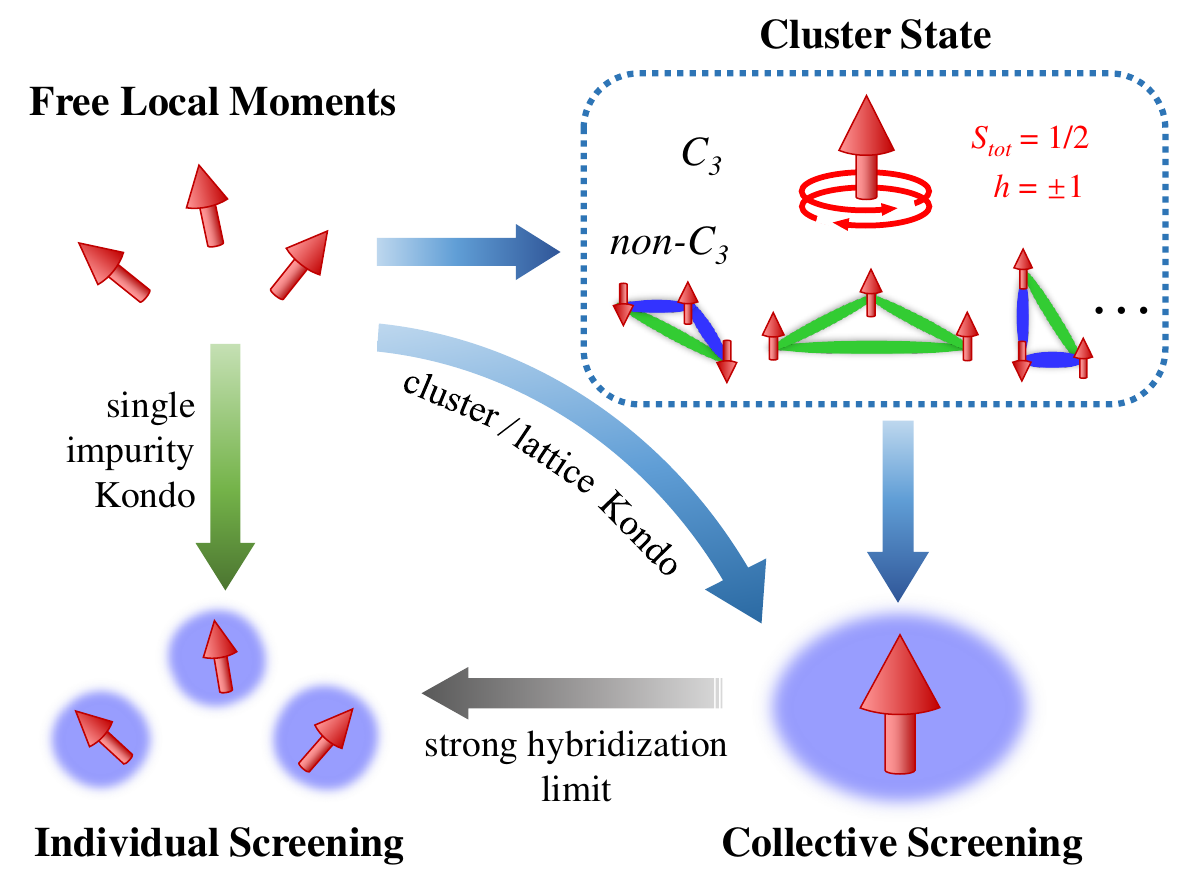}
		\par\end{centering}
	\caption{Schematic plot of the multi-impurity states evolving successively from free local moments, to cluster spin states, to collective Kondo screening for moderate hybridization, and to individual screening at strong hybridization limit.}
	\label{fig4} 
\end{figure}

Our method can be readily extended to general $N$-impurity models with more complicated baths \cite{Liu2016PRB,Wojcik2023PRB,Lotem2023PRB}. However, computations for larger $N$ or $N_A$ require more efficient NRG algorithms, which will stimulate future code development. Our results also have important implications on the Kondo lattice physics. The impurities always tend to form some collective degrees of freedom to be successively screened with lowering temperature or increasing hybridization. As the number of impurities grows, the number of collective degrees of freedom also increases, causing more plateaus and eventually continuous depletion of the magnetic entropy as observed in real materials. Therefore, the screening in multi-impurity or lattice Kondo systems with shared baths is generally nonlocal except for extremely large Kondo couplings. This reveals a fundamental aspect of the Kondo lattice physics beyond the conventional Doniach paradigm \cite{Yang2008Nature,Lonzarich2017RPP, Wang2021PRB,Wang2022PRB,Yang2022}.

This work was supported by the National Key R\&D Program of China (Grants No. 2024YFA1408602 and No. 2022YFA1402203), the National Natural Science Foundation of China (Grants No. 12174429, No.12304174, No. 12547101, and No. 12404168),  and the Strategic Priority Research Program of the Chinese Academy of Sciences (Grant No. XDB33010100). 

D.H. and J.W. contributed equally to this work.


\begin{thebibliography}{99}
\bibitem{units} For simplicity, the $g$ factor, the Bohr magneton $\mu_B$, and the Boltzmann's constant $k_B$ are all set to unity.
\end{thebibliography}
\end{document}